\documentclass[showpacs,prl,twocolumn,amsmath,amssymb,superscriptaddress,amssymb]{revtex4}
\usepackage{amsfonts}
\usepackage{amsmath}
\usepackage{amssymb}
\usepackage{CJK}
\usepackage{graphicx}%
\providecommand{\U}[1]{\protect \rule{.1in}{.1in}}

\begin{document}
\title{Magneto--optical matter wave Bragg diffraction }

\title{Bragg diffraction of a matter wave driven by a pulsed non-uniform magnetic field }
\author{Yueyang Zhai}
\affiliation{School of Electronics Engineering and Computer Science, Peking University, Beijing 100871, China}

\author{Peng Zhang }
\affiliation{School of Electronics Engineering and Computer Science, Peking University, Beijing 100871, China}
\author{Xuzong Chen }
\affiliation{School of Electronics Engineering and Computer Science, Peking University, Beijing 100871, China}
\author{Guangjiong Dong}\email{dong.guangjiong@gmail.com}
\affiliation{State Key Laboratory of Precision Spectroscopy, East China Normal University,
3663 North Zhongshan Road, Shanghai, 200062, China}
\author{Xiaoji Zhou }\email{xjzhou@pku.edu.cn}
\affiliation{School of Electronics Engineering and Computer Science, Peking University, Beijing 100871, China}

\pacs{03.75.-b,42.25.Fx,67.85.Hj}

\begin{abstract}
We have performed a principle-proof-experiment of a magneto-optical
diffraction (MOD) technique that requires no energy level splitting
by homogeneous magnetic field and a circularly polarized optical
lattice, avoiding system errors in an interferometer based on the
MOD. The principle for this new MOD is that asynchronized switching
of quadrupole trap and Ioffe trap in a
quadrupole-Ioffe-configuration trap can generate a residual magnetic
force to drive a Bose-Einstein condensate (BEC) to move. We have
observed asymmetric atomic diffraction resulting from the asymmetric
distribution of the Bloch eigenstates involved in the diffraction
process when the condensate is driven by such a force, and
matter-wave self-imaging due to coherent population oscillation of
the dominantly occupied Bloch eigenstates. We have classified the
mechanisms that lead to symmetric or asymmetric diffraction, and
found that our experiment presents a magnetic alternative to a
moving optical lattice, with a great potential to achieve a very
large momentum transfer ($>110 \hbar k$) to a BEC using
well-developed magnetic trapping techniques.


\end{abstract}
\maketitle

Atomic/molecular matter waves \cite{pi,AP} have played an important role in
fundamental research and many practical applications, such as atomic clocks
\cite{clock, Ludlow}, gravitational-wave detection \cite{sav},
gravito-inertial sensors \cite{ch,gradiometers}, atom lithography
\cite{mutzel, JPD}, rotation sensing \cite{wu0,wu}, detection of tiny effects
of general relativity \cite{Savas,mu}, measurement of atom surface
interactions \cite{as}, generation of quantum correlated atom pairs
\cite{kh1}, and dispersion manipulation \cite{Murch}. Among these studies, coherent splitting of atomic beams is a key
technique \cite{pi,pra Splitting,interferomers2,mo3}. Matter wave Bragg diffraction, analogous to
diffraction of an optical beam by a periodic medium, has been intensively
investigated for splitting a matter wave into a superposition of momentum
states using an optical lattice \cite{ear1,HolgerLMS,ear2} or a magnetic
lattice \cite{ma,ma1,ma2}. High-momentum transfer splitters are
essential for high-precision atom interferometers \cite{li}. So, there is a
continuous endeavor \cite{102,interferomers2,Kovachy} to increase the
momentum transfer to the BEC. Recently, momentum transfer of 102$\hbar
k_{L}$ from optical fields to a condensate has been successfully
demonstrated \cite{102}.

A moving optical lattice \cite{Morsch,Fallani1,Fallani2,Mun}, formed by two
counterpropagating optical fields with unequal frequencies \cite{dong03}, has
shown promise for mater wave splitting, because the lattice velocity $v_{L}$
opens a new degree of freedom for controlling the momentum transfer, i.e., the
diffracted atoms are prepared to populate in $\pm v_{L}$ under the resonant
Bragg scattering condition \cite{Holger} in the reference frame where the lattice is stationary, provided that $v_L$ equals to the single-photon recoil velocity $\hbar k_L/m$. More interestingly, a large lattice
velocity is capable of realizing large angle beam splitting, as demonstrated
by a recent experiment \cite{Clad,HolgerLMS}. Furthermore, the approach can
also realize asymmetric atomic diffraction (the split beams have unequal
population) \cite{Holger}, which could be useful for controlling population
ratios for the two split beams. In contrast, when an atomic gas at rest is
probed by a static optical lattice, no asymmetric atomic diffraction could be
induced in general. Recently, however, one experiment \cite{deng 2008} shows
that using two counterpropagating optical fields with equal frequency but
unequal intensities allows asymmetric atomic diffraction for an atomic gas
initially at rest. This counter-intuitive result is explained later by the
local field effect (LFE) \cite{dong}, i.e., an asymmetric optical lattice, due
to asymmetric scattering of the incident optical beams by the condensate,
generates the asymmetric diffraction.

Currently, most of matter wave diffraction experiments apply either an optical
lattice or a magnetic lattice. \textit{Meanwhile}, magneto-optical diffraction
of atoms in a magnetic field by a circularly polarized optical standing wave
\cite{mo2} has also been investigated for static atomic clouds to achieve a
large momentum transfer. In this approach, Zeeman splitting of energy levels in an external magnetic field is used  such that the energy difference of the atoms and the considerably induced
wavefront curvature could lead to a systematic phase error
in an interferometer based on the splitter \cite{pra Splitting}.

In this paper, we experimentally demonstrate a magneto-optical atomic
diffraction, combining a quadrupole-Ioffe-configuration (QUIC) trap and an
optical standing wave, which is not circularly polarized as in \cite{mo2}. We observe an asymmetric atomic
diffraction and matter wave self-imaging, when the quadrupole trap and the
Ioffe trap are switched off asynchronizedly, generating a residual magnetic
force and driving the condensate to move. Such an asymmetric atomic diffraction is due to the non-zero initial velocity of the condensate from the magnetic acceleration, rather than LFE. Our
experiment shows the potential of a magnetic force in optical atomic
diffraction. Magnetic driving of a BEC in this way can be an alternative to a
moving optical lattice for achieving a momentum transfer to a BEC.

Our diffraction experiment is performed as follows. A cigar shaped BEC of
$N$=$1\times10^{5}$ $^{87}$Rb atoms in $\left\vert F=2,m=2\right\rangle $
state with a longitudinal Thomas-Fermi radii $40\mu$m and a transverse radii $4\mu$m is
first prepared in an anisotropic QUIC trap with the axial frequency 20 Hz and
the radial frequency 220 Hz \cite{pra xiongwei, liuxinxing,zhou}. A one dimensional
optical standing wave with a wavelength 852nm is incident onto the condensate
along the long axis direction ($x$ axis). In our experiment, the standing wave is formed with
a retroreflected laser beam, well focused on the BEC to a waist of 110 $\mu$m.
The incident optical field is a square light pulse with an intensity $I_{1}=1.03\times10^{5}
$ $mW/cm^{2}$.
But the reflected light $\mathit{I}_{2}$ is controlled by a tunable light
attenuator which is put in front of the reflection mirror. When the QUIC trap
is switched off, the optical standing wave is not turned on immediately, but
is delayed for a time $\Delta T$ for sufficiently exploiting the residual
magnetic force. After switching-off of the optical fields for 28 ms, an
absorption image is taken.

\begin{figure}[h]
\begin{center}
\scalebox{0.43}[0.45] {\includegraphics*
[60pt,290pt][590pt,800pt]{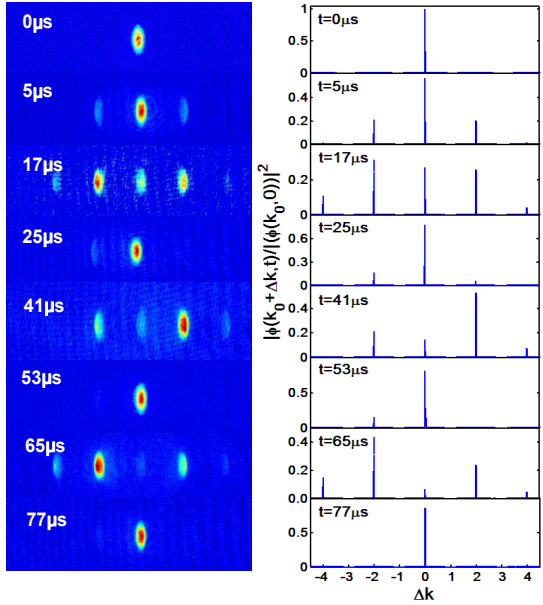}}
\end{center}
\caption{(Color online) The left panel shows the absorption images of the diffracted BEC for
different pulse durations. The right panel shows the related simulation
results of the condensate momentum distribution ($\hbar \Delta k$ is the dimensionless net momentum each atom obtains in each diffraction order).}%
\label{result}%
\end{figure}

The left panel of Fig. \ref{result} shows the absorption images for different
pulse durations and a fixed delay time $\Delta T=400$ $\mu$s with
$I_{2}=0.8I_{1}$. Asymmetric diffraction is one prominent feature of our
experiment, and self-imaging of the matter wave occurs roughly at 25 $\mu$s,
53 $\mu$s and 77 $\mu s$. These experimental phenomena remind us of a recent
experiment \cite{deng 2008} dominated by LFE \cite{dong}, which is also our
original motivation to study this Bragg diffraction. However, simulation of
our experiment with the LFE fails in fitting our experimental phenomena with
an initial wave function $\psi(x,0)=C\exp[-x^{2}/(2w^{2})]$ ($C$=$(\sqrt{\pi
}w)^{-1/2}$) for a BEC released from a harmonic trap, where $w$ is the full
width at half maximum; because the huge detuning of about 10$^{15}$ Hz in our
experiment is six orders of magnitude larger than that in Ref. \cite{deng
2008, dong}, such that the local refraction index, which is inversely
proportional to the detuning, has little influence on the propagation of the
optical fields. Neglecting the local-field effect, the optical lattice
potential is proportional to $\sqrt{I_{1}I_{2}}$. Thus, for fixed $I_{1}I_{2}%
$, the diffraction processing should be the same for different values of
$I_{1}$ and $I_{2}$.

In our experiment, before the optical field is switched on, the condensate is
accelerated during the switching-off of the magnetic trap. The QUIC trap
generated by the driven currents cannot be switched off instantly but with a
relaxation, such that a residual magnetic force is produced to drive the BEC
into motion. The force at position $X$ is given by
\begin{equation}
F(X,\!T)\!\!=\!-\mu\!\! \left[  \! i_{I} \frac{\partial}{\partial X}%
\!f_{I}(\!X\!)e^{-T/\tau_{I}}\!\!+\!i_{Q}\frac{\partial}{\partial X}%
f_{Q}(\!X\!)e^{-T/\tau_{Q}}\!\right]  , \label{force}%
\end{equation}
where $f_{I}$ ($f_{Q}$), $i_{I}$($i_{Q}$), and $\tau_{I}$ ($\tau_{Q}$) are
respectively the structure function, driven currents, and relaxation time
of\ the Ioffe-type (quadrupole) trap \cite{Xiuquan}, $\mu$ is the
atomic magnetic moment. At $T$=0, $F$=0. When $\tau_{I}\neq\tau_{Q}$, \ a
nonzero magnetic force $F(X,T)$ drives the BEC to move, experimentally
confirmed by absorption images (not shown here) of the BEC during releasing process.

We now show the condensate motion is a new mechanism beyond LFE for asymmetric
diffraction. When the lattice with a large detuning is not distorted by the
LFE \cite{dong}, the matter wave function $\psi(x,t)$ satisfies the Mathieu
equation,
\begin{equation}
i\frac{\partial}{\partial t}\psi(x,t)=[-\frac{\partial^{2}}{\partial x^{2}%
}+2q\cos\left(  2x\right)  ]\psi(x,t), \label{ma}%
\end{equation}
Here, dimensionless $t$, $x$ and $q$ are respectively related to the real time
$T$, position $X$ and dipole potential $V_{0}$, according to $t=\omega_{r}T$,
$x=k_{L}X$, $q=V_{0}/(2\hbar\omega_{r})$ with the wave vector of the pump
fields $k_{L}$ and the recoil frequency $\omega_{r}=\hbar k_{L}^{2}/(2m)$. Eq.
(\ref{ma}) has Bloch eigenfunctions corresponding to eigenenergy
$\epsilon_{2N+s}$ as \cite{Shirts},
\begin{equation}
\varphi_{N,s}=\exp[i(2N+s)x]%
{\textstyle\sum\limits_{n}}
c_{2n}^{2N+s}\exp(i2nx), \label{ei}%
\end{equation}
in which -1$\leqslant s\leqslant1$, and $qc_{2n-2}^{2N+s}+(s+2N+2n)^{2}%
c_{2n}^{2N+s}+qc_{2n+2}^{2N+s}=\epsilon_{2N+s}c_{2n}^{2N+s}$. The Mathieu
function has properties \cite{Shirts}: (1) $\epsilon_{2N+s}=\epsilon_{-2N-s}$;
(2) $c_{2n}^{2N+s}=c_{-2n}^{-2N-s}$.

In the momentum space, the wave function of BEC is $\phi(k,t)=%
{\textstyle\int\limits_{-\infty}^{\infty}}
\psi(x,t)\exp(ikx)dx$. For a moving condensate with the initial wave function
$\psi(x,0)=C\exp[-x^{2}/\left(  2w^{2}\right)  ]\exp(-ik_{0}x)$, using the
eigenvalues of the Mathieu equation, we obtain $\phi(k,t)=\!\!\!\!
{\textstyle \sum\limits_{\{N,m,n\geq-N-\frac{1+k}{2}\}}^{\{n\leq-N-\frac
{1-k}{2}\}}} 2\pi C\times c_{2m}^{-k-2n}c_{2n}^{-k-2n}e^{-\frac{w^{2}}%
{2}(k+2n-2m-k_{0})^{2}}e^{-i\varepsilon_{-k-2n}t}$. For a large width $w$,
using the steepest descent approximation, the atoms are populated in momentum
space around $k=k_{0}+2j$ ($j$ is an integer). Denoting $\phi_{j}^{k_{0}%
}\equiv$ $\phi(k_{0}+2j,t)$, we have
\begin{equation}
\phi_{j}^{k_{0}}\approx\mathcal{N}e^{-i\varepsilon_{-k_{0}}t}%
{\textstyle\sum\limits_{N}}
{\textstyle\sum\limits_{n\geq-N-\frac{1+k}{2}}^{n\leq-N-\frac{1-k}{2}}}
c_{2K}^{-k_{0}-2K}c_{2K-2j}^{-k_{0}-2K}e^{-i\Delta\varepsilon_{-2K}t},
\label{m3}%
\end{equation}
in which $K=j+n$, $\Delta\varepsilon_{-2K}=\varepsilon_{-k_{0}-2K}%
-\varepsilon_{-k_{0}}$, and $\mathcal{N}$ is a normalization factor (in the
followed part, all irrelevant constants are absorbed in this factor).

\begin{figure}[th]
\begin{center}
\scalebox{0.40}[0.38] {\includegraphics*
[55pt,380pt][595pt,830pt]{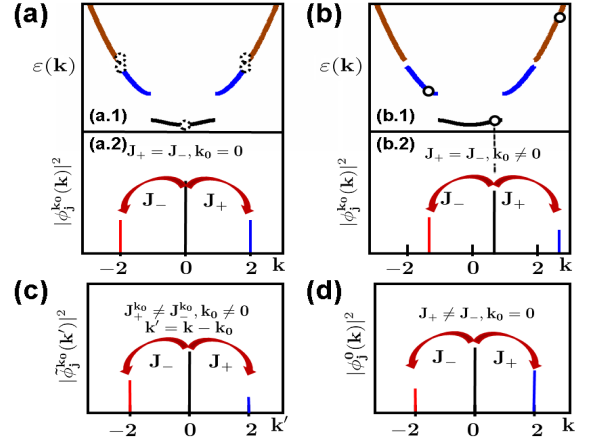}}
\end{center}
\caption{(Color online) Different mechanisms for symmetric and asymmetric atomic diffraction.
(a) and (b) are respectively for diffractions of a static and moving BEC in
which the involved Bloch eigenstates are shown in circles in (a.1) and (b.1),
$\epsilon(k)$ is the Bloch eigen-energy. $\phi_{j}^{\mathbf{k}_{0}}$ in (a.2)
or (b.2) is the wave function of the diffracted BEC in momentum space. (c)
Diffraction of atoms in the frame of the moving condensate, with $\tilde{\phi
}_{j}^{\mathbf{k}_{0}}$ the atomic wave function in this frame. (d) Atomic
diffraction with the local field effect, with $\phi_{j}^{0}$ the wave function
of a stationary BEC. $k$ is dimensionless vector in the reciprocal space.}
\label{m}%
\end{figure}

Eq.~(\ref{m3}) shows that eigenstates corresponding to eigenenergy
$\varepsilon_{-k_{0}-2K}$ have been involved in diffraction process. Figs.
\ref{m}(a.1) and (b.1) schematically show the involved eigenstates
respectively for $k_{0}=0$ and $k_{0}\neq0$ in an extended band structure of
the Mathieu equation. When $k_{0}=0$ (Fig.~\ref{m}(a.1)), the involved
eigenstates shown in dotted circles are symmetrically distributed, thus, we
have symmetric diffraction, i.e. $\phi_{j}^{k_{0}}=\phi_{-j}^{k_{0}}$. When
$k_{0}\neq0$ (Fig.~\ref{m}(b.1)), the involved eigenstates shown in solid
circles are not symmetrically distributed, such that $\phi_{j}^{k_{0}}$ does
not equal to $\phi_{-j}^{k_{0}}$ in general. We numerically calculate momentum
spectrum using the experimental parameters, $k_{0}=0.32$ and $V_{0}%
=4.45\hbar\omega_{r}$. $k_{0}$ is obtained by measuring the motion of the
condensate peak. \textbf{ }Actually, assuming that the motion is driven by the
residual magnetic force of Eq.~(\ref{force}) with the experimental parameters
($\tau_{I}=66\mu s$, $\tau_{Q}=87\mu s$, $i_{0}=21$A), the simulated value of
$k_{0}\cong0.3$ is consistent with the measured value. The numerical results
are shown at the right panel of Fig.~\ref{result}. Comparison of the left and
right panels of Fig.~\ref{result} shows a good agreement of the theory with
the experimental results, indicating that the BEC indeed obtains a nonzero
velocity due to asychronized switching-off of the quadrupole trap and Ioffe trap.

Finally, we present a classification of all observed symmetric or asymmetric
atomic diffraction phenomena. When the initial matter wave has a narrow
momentum distribution, no matter the local field effect is involved or not,
$\phi_{j}^{k_{0}}$ satisfies%
\begin{equation}
i\frac{\partial}{\partial t}\phi_{j}^{k_{0}}=(k_{0}+2j)^{2}\phi_{j}^{k_{0}%
}+J_{-}\phi_{j-2}^{k_{0}}+J_{+}\phi_{j+2}^{k_{0}}. \label{me}%
\end{equation}
where $J_{\pm}$ is transition rate from the momentum component $\phi
_{j}^{k_{0}}$ to $\phi_{j\pm2m}^{k_{0}}$. \ When the local field effect is
negligibly small, $J_{-}=J_{+}=q$ is not dependent on $k_{0}$. However, the
symmetric hopping cannot guarantee symmetric diffraction. When $k_{0}=0$,
there is symmetric distribution (Fig. \ref{m}(a.2)). When $k_{0}\neq0$, the
\textit{j}-th order diffraction component is related to the momentum
$k_{0}+2j$ (Fig. \ref{m}(b.2)), and the matter wave diffraction is asymmetric.

To see how the asymmetric diffraction happens, we turn to the reference frame
moving at the speed $k_{0}$, in which the diffraction equation of the
wavefunction $\tilde{\phi}_{j}^{k_{0}}$ for the \textit{j}-th order diffracted atoms is
$i\frac{\partial}{\partial t}\tilde{\phi}_{j}^{k_{0}}=4j^{2}\tilde{\phi}%
_{j}^{k_{0}}+J_{-}^{k_{0}}\tilde{\phi}_{j-2}^{k_{0}}+J_{+}^{k_{0}}\tilde{\phi
}_{j+2}^{k_{0}},$ with $J_{\pm}^{k0}=qe^{\pm i2k_{0}t}$. In this moving frame,
the \textit{j}-th order diffraction corresponds to 2$j$ momentum as in the
static frame; however, $J_{+}^{k_{0}}\neq J_{-}^{k_{0}}$, leading to
asymmetric diffraction (Fig. \ref{m}(c)). This mechanism is different from the
mechanism by LFE with the incident and counterpropagating lights of unequal
intensities, as shown in Fig. \ref{m}(d), where $J_{-}$ and $J_{+}$ are time
dependent and are not equal ($J_{-}(t)\neq J_{+}(t)$). In the LFE-dominant case,  the condensate leads to unequal scattering of the incident counter-propagating optical fields with unequal intensities,such that the spatial inversion symmetry of the macroscopic wavefucntion for the condensate is induced, and consequently the asymmetric momentum transfer occurs.

We have to emphasize that the asymmetry of the atomic diffraction is induced by the initial acceleration of the BEC.
The asymmetric atomic diffraction mechanism in Fig.~\ref{m}(b)-(c) also
applies to that with a moving optical lattice \cite{ku}. Thus, our experiment
presents a magnetic alternative to the latter in transferring momentum to the condensate.Driving a static BEC to a very
high momentum with a moving optical lattice may require a strong laser
intensity \cite{ko}. Our experiment shows that we also could exploit the
well-established magnetic trap techniques to achieve a very high momentum
transfer to a BEC, by reducing the relaxation time of the Ioffe trap and
increasing the quadrupole current. We have performed a theoretical calculation
of the acceleration got a BEC with QUIC trap parameters $i_{Q}=i_{I}=70A$,
$\tau_{Q}=5ms$, and $\tau_{I}=50\mu s$. Figure 3(a) displays the spatial
distribution of the residue magnetic field at different moments after the QUIC
trap currents are switched off.  When Ioffe trap is rapidly switched off
($\tau_{I}\ll\tau_{Q}$), the gradient magnetic field forming the quadrupole
trap coherently drives the BEC. Figure 3(b) shows the related condensate
velocity. With this specially designed QUIC trap,  the momentum of the
condensate can reach up to $110\hbar k_{L}$ within $3ms$. With our experimental setup, we have been able to increase momentum of the BEC to more than $2\hbar k$ by using a current of 24A in QUIC trap. For further increasing the momentum of the condensate, a better water-cooling system is needed. However, the numerical simulation based on our current experimental configuration indicates that the condensate velocity can be accelerated to more than $100\hbar k$ when a current of 70A is given. Such big current has already been achieved in some experiments \cite{C_Chin}. Our work could pave a way for studying the diffraction theory \cite{Holger} of a fast-moving BEC to achieve a large-momentum transfer atom interferometer.
\begin{figure}[ptb]
\centering
\includegraphics[height=8cm]{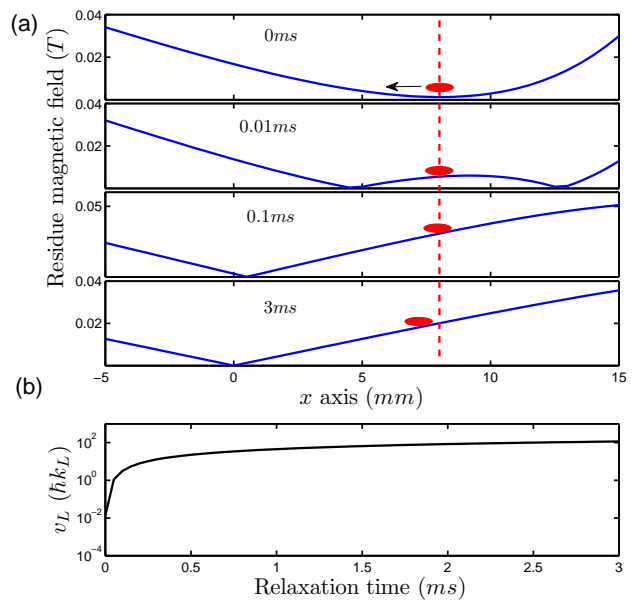}
\caption{(Color online)(a) Spatial distribution of the magnetic field within the QUIC trap at the different
moments after the QUIC trap is switched-off. The condensate is plotted to
display the coherent driving. (b) The time evolution of the condensate
velocity due to the residue magnetic field after the QUIC trap is switched off.
}%
\end{figure}

Our theory also predicts that, when the strength of one of the pump field
is lowered down, i.e., the optical lattice is shallower, the asymmetry of the atomic diffraction would be enhanced.
This prediction can be tracked by expanding the light-atom coupling Hamiltonian with the
shifted plane wave modes. In this presentation, the effective Hamiltonian
reads, $H=\sum_{i}E_{i} \vert \phi_{k_{0}}^{i} \rangle \langle
\phi_{k_{0}}^{i} \vert +J\sum_{i} \{   \vert \phi_{i}^{k_{0}%
}\rangle  \langle \phi_{i+1}^{k_{0}} \vert + \vert
\phi_{i+1}^{k_{0}} \rangle  \langle \phi_{i}^{k_{0}} \vert
 \}$, where, $E_{i}=\frac{\left(  \hbar k_{0}+2i\hbar k\right)  ^{2}%
}{2M}$ is the energy level of the free atomic system without lattice. Due to the acceleration of the
condensate, the single particle energy of the quasi mode is asymmetric about
the central mode, thus the effective detuning of off diagonal Rabi frequency
is asymmetric, resulting an asymmetric atomic Bragg diffraction patterns.
However, in the large coupling limit, where the kinetic energy of the atoms is
vanishingly small compared to the light-atom coupling, the free part of the
single particle energy can be neglected, rendering the Hamiltonian taking a
symmetric form. Therefore, the rabi oscillations toward the forward backward
directions are nealy identical, leading to more\ symmetric matter-wave
diffractions in the time-domain. While in the other limit where the lattice is shallower, the asymmetry of the atomic diffraction would be more prominent.

Eq.~(\ref{m3}) has also been used to explain the matter wave self-imaging as
atomic center-of-mass motion induced interference \cite{deng 2008,dong,Ovchinnikov}. This
does not present an analytical result for the self-imaging time. However,
Eq.~(\ref{m3}) could be used to give a good estimation of the self-imaging
time. Not all eigenstates of eigenenergy $\varepsilon_{-k_{0}-2K}$ are
essentially involved, so we have a cutoff $N_{\max}$ for $N$. Thus when
$t=T_{\text{si}}$ where $T_{\text{si}}$ is the least common multiple of the
periods of all essentially involved eigenstates, $\phi_{0}^{k_{0}%
}(T_{\text{si}})\approx$ $\phi_{0}^{k_{0}}(0)$, \ i.e., matter wave
self-imaging occurs. For example, in Fig.~\ref{result}, $\left\vert \phi
_{j=0}^{k_{0}}(t)\right\vert ^{2}\approx\mathcal{N}\left\vert (c_{0}%
^{k_{0}\newline})^{2}+(c_{2}^{-2-k_{0}\newline})^{2}e^{-i\Delta\varepsilon
_{-2}t}+(c_{2}^{-2+k_{0}\newline})^{2}e^{-i\Delta\varepsilon_{-2}t}\right\vert
^{2}$ in which $(c_{0}^{k_{0}\newline})^{2}=0.588$, $(c_{2}^{-2-k_{0}\newline%
})^{2}=0.335$, $(c_{2}^{-2+k_{0}\newline})^{2}=0.052$, $\Delta\varepsilon
_{2}=7.337$, and $\Delta\varepsilon_{-2}=12.419$. Since $(c_{2}^{-2+k_{0}%
\newline})^{2}\ll(c_{2}^{-2-k_{0}\newline})^{2}$, $\Delta\varepsilon_{-2}$ is
the dominant frequency. The self-imaging time $T_{\text{si}}$ is roughly given
as $T_{\text{si}}\approx2n\pi/(\Delta\varepsilon_{-2}\omega_{r})=25.5n$ $\mu$s
($n=1,2,3,...$), which agrees well with the experimental values. Thus, the
matter-wave self-imaging is a kind-of coherent-population oscillation between
two Bloch states $\epsilon_{-k_{0}}$ and $\epsilon_{-k_{0}-2}$.

In summary, we have performed experimental study of diffraction of a BEC
released from a QUIC trap by an optical standing wave and observed asymmetric
diffraction and matter wave self-imaging. In contrast with \cite{deng 2008,dong},
the lattice is not distorted in our experiment. Thus, the experimental
phenomena is induced by a new mechanism beyond the local field effect. We
find that the BEC obtains a velocity due to a residual magnetic force during
the asynchronized switching-off of the quadrupole trap and Ioffe trap, before
the optical lattice is switched on. The initial velocity leads to asymmetric
distribution of the involved Bloch eigenstates in momentum space, such that
asymmetric diffraction occurs. The matter wave self-imaging is
analytically explained as a coherent-population oscillation between two Bloch
eigenstates. Finally, we have presented a clarification of the mechanisms that
leads to symmetric or asymmetric diffraction.

Compared to other diffraction schemes using magneto-optical potential
\cite{mo2}, our experiment using atoms in the ground state applies no
circularly polarized optical lattice, thus the new magneto-optical diffraction
technique can avoid the phase error in an interferometer due to energy
difference of atoms at different energy levels \cite{pra Splitting}. Moreover, in our approach, the magnetic acceleration and optical
diffraction is separated. Therefore, this approach is free from the non-uniform and
fluctuation of a magnetic field and the corresponding energy splitting due to
Zeeman effect. In the metrology experiments with lattice-based matter-wave-accelerations, the unit of momentum transfer is $2 \hbar k$ \cite{102}. This method has been proved to be able to achieve a very large momentum transfer. Our approach is an alternative method based on gradient magnetic field for accelerating the atoms. Furthermore, in \cite{102} the wave front distortions of light pulse broadened the momentum and achieved a contrast of $18\%$ with $102\hbar k$ beam splitters. By using gradient magnetic field to accelerate atoms, the achieved momentum distribution is as narrow as that of the original condensate which leads to a high contrast. In our experiment, the current in the magnetic coils can be maintained with stability on the order of $10^{-4}$, and the length of $\Delta T$ can be controlled with precision on the order of several tens of nanoseconds. Then by fixing the timing sequence of our experiment, the momentum that transferred to condensate can be controlled and distinguished with precision $0.01\hbar {k_L}$ according to the absorption images. Therefore, our method shows potential in metrology field.

To achieve a high momentum transfer, it is hard to avoid asymmetric
momentum splitting of a condensate through diffraction \cite{note}. This
asymmetric diffraction could be used to develop intensity-imbalanced matter
wave interferometers, analogous to intensity-imbalanced optical
interferometers£¬which have been used for monitoring the beam size in a
particle accelerator \cite{j}, or for reducing back action in a two-path
interferometer \cite{ba}. It is worthy of exploiting the asymmetric
diffraction for precision measurement with matter waves in future. Finally, the
asymmetric splitting of matter waves could be used to study symmetry-broken
spontaneous four-wave mixing with matter waves \cite{four}, to generate
directional correlated atom pairs.

This work was supported by the National Basic Research Program of China under
Grants No. 2011CB921604 and 2011CB921501, the National Natural Science
Foundation of China under Grants No. 10874045, No. 10828408, No. 11334001 and No. 61078026, RFDP No.20120001110091.
G.J. Dong acknowledge the support of the Program of Introducing Talents of
Discipline to Universities (B12024).

\end{document}